\documentclass[aps,11pt]{revtex4}
\usepackage{dcolumn}
\usepackage{graphicx}
\usepackage{amsmath}
\usepackage{amsfonts}
\usepackage{amssymb}
\usepackage{psfrag}
\usepackage{wrapfig}
\usepackage{subfigure}
\usepackage{makeidx}
\usepackage{bm}
\usepackage{hyperref}
\usepackage{color}
\makeatletter

\newcommand{\Rmnum}[1]{\expandafter\@slowromancap\romannumeral #1@}

\makeatother

\begin{document}
\title{Triple points and phase diagrams of Born-Infeld AdS black holes in 4D Einstein-Gauss-Bonnet gravity}

\author{Chao-Ming Zhang$^{1}$\footnote{843395448@qq.com}, De-Cheng Zou$^{1}$\footnote{ Corresponding author:dczou@yzu.edu.cn}
and Ming Zhang$^{2}$\footnote{zhangming@xaau.edu.cn}}

\address{$^{1}$Center for Gravitation and Cosmology, College of Physical Science and Technology, Yangzhou University, Yangzhou 225009, China\\
$^{2}$Faculty of Science, Xi'an Aeronautical University, Xi'an 710077 China}

\date{\today}

\begin{abstract}
\indent

By treating the cosmological constant as a thermodynamic pressure,
we investigate the thermodynamic behaviors of Born-Infeld AdS black hole in
4D Einstein-Gauss-Bonnet (EGB) gravity.
The result shows that the Van der Waals like small/large black hole (SBH/LBH) phase
transition always appears for any positive parameters $\alpha$ and $\beta$.
Moreover, we observe a new phenomenon of small/intermediate/large black
hole (SBH/IBH/LBH) phase transition with one tricritical and two critical points
in the available parameter region for $\alpha$ and $\beta$.
This behavior is reminiscent of the solid/liquid/gas phase transition.
\end{abstract}


\maketitle

\section{Introduction}

Due to the AdS/CFT correspondence, lots of attention have been attracted to study thermodynamics and
phase transition of black holes in anti-de Sitter(AdS) space during past decades\cite{Maldacena:1997re,Gubser:1998bc,Witten:1998qj}.  Particularly interesting is the Hawking-Page transition\cite{Hawking:1982dh}
which corresponds to the confinement/deconfinement phase transition in the dual quark gluon plasma \cite{Witten:1998zw}.
Recently thermodynamics of AdS black holes has been generalized to the extended phase space, where the
cosmological constant is treated as the pressure of the black hole $P=-\frac{\Lambda}{8\pi}$ \cite{Dolan:2011xt,Dolan:2010ha}.
Later, Ref.\cite{Kubiznak:2012wp} recovered that a first order small black hole and large black hole (SBH/LBH) phase
transition for four dimensional Reissner-Nordstr\"{o}m (RN) AdS black hole is superficially analogous
to a liquid-gas phase transition of the Van der Waals fluid, and both systems share the same critical
exponents near the critical point. This analogy has been also
generalized to different charged black holes and rotating black holes in AdS space in the extended phase space \cite{Gunasekaran:2012dq,Hendi:2012um,Hendi:2015hgg,Zhao:2013oza,Dehghani:2014caa,Hennigar:2015esa,Zhang:2014jfa,
Rajagopal:2014ewa,Altamirano:2014tva,Sadeghi:2016dvc,Cheng:2016bpx,Wei:2015ana,Hendi:2015cqz,Mo:2016ndm,Xu:2014kwa}.

In the low energy limit, string theories give rise to effective models of gravity in higher dimensions
that involve higher order curvature terms in the action.
Among the higher curvature correction gravities, the most
extensively studied theory is the so-called Gauss-Bonnet(GB) gravity\cite{Boulware:1985wk}.
However, in four dimensions the GB term is a total derivative,
and has no influence on the field equation. It's interesting to note that
some authors considered the GB gravity by rescaling the GB coupling parameter
$\alpha \to \alpha/(D-4)$ in four dimensions \cite{Tomozawa:2011gp,Cognola:2013fva},
and then the GB term gives rise to non-trivial dynamics. Taking the similar technique,
Glavan and Lin \cite{Glavan:2019inb} presented 4D Einstein Gauss-Bonnet gravity(4D EGB).
Its main feature is that the static spherically symmetric solution
of this theory is free from the singularity problem for a positive GB coupling constant.
Later, the solutions of Maxwell charged GB AdS black hole \cite{Fernandes:2020rpa},
Hayward AdS \cite{Kumar:2020xvu} and Bardeen AdS \cite{Kumar:2020uyz} black holes
have been also constructed in the 4D EGB gravity. Moreover, the thermodynamics and phase transition
of Maxwell charged GB AdS black hole \cite{Hegde:2020xlv,Hegde:2020yrd,Wei:2020poh,HosseiniMansoori:2020yfj},
nonlinear electrodynamics charged GB AdS black hole \cite{Singh:2020mty}
and Bardeen AdS \cite{Singh:2020xju} black hole further have been also
discussed in the extended phase space.

As well known, a point-like charge in Maxwell's electromagnetic field theory usually brings about
infinite self-energy, since it is allowed a singularity at the charge position.
Then, Born, Infeld \cite{Born:1934gh} and Hoffmann \cite{Hoffmann:1935ty} introduced Born-Infeld(BI)
electromagnetic field to overcome infinite self-energy problem by imposing a maximum strength
of the electromagnetic field. Moreover, BI type effective action arises in an open superstring
theory and D-branes are free of physical singularities.
In recent years, the solution and related thermodynamic properties of BI black hole have
received some attentions \cite{Dey:2004yt,Cai:2004eh}. Particularly, in the extended space,
the BI AdS black hole exhibits an interesting reentrant phase transition in four dimensional
Einstein-Born-Infeld(EBI) gravity, besides the usual Van der Waals liquid-gas like SBH/LBH phase
transition \cite{Gunasekaran:2012dq,Zou:2013owa}. Recently, Ref.\cite{Yang:2020jno} proposed
a static and spherically symmetric BI AdS black hole solution in the novel 4D EGB gravity,
and analyzed some basic thermodynamics of the BI AdS black hole.
In this paper, we will generalize the discussions to consider the possible phase
transition and critical phenomena for BI AdS black holes in the extended phase space.

This paper is organized as follows. In Sec.~\ref{2s}, we reconsider the BI AdS black hole
solution and its thermodynamics in the 4D EGB gravity. In Sec.~\ref{3s},
we will investigate phase transition and critical behavior of
BI AdS black holes in the 4D EGB gravity. We will summarize our results in Sec.~\ref{4s}.

\section{Solution and thermodynamics of Born-Infeld AdS black hole }
\label{2s}

The action of $D$-dimensional EGB gravity  minimally coupled to BI electrodynamics field
in the presence of a negative cosmological constant is given by \cite{Yang:2020jno}
\begin{eqnarray}
{\cal I}&=&\frac{1}{16\pi}\int d^{D}x \sqrt{-g}\left(R-2\Lambda+\frac{\alpha}{D-4}{\cal G}+{\cal L_{BI}} \right), \label{action}
\end{eqnarray}
where  $\Lambda\equiv -\frac{(D-1)(D-2)}{2l^2}$ is a negative cosmological constant, ${\cal G}$ is the Gauss-Bonnet term $R^2-4R_{\mu\nu}R^{\mu\nu}+R_{\mu\nu\rho\sigma}R^{\mu\nu\rho\sigma}$, $\alpha$ is the Gauss-Bonnet coefficient,
and ${\cal L_{BI}}$ is the Lagrangian of the BI electrodynamics with
\begin{eqnarray}
4\beta^2\left(1-\sqrt{1+\frac{F_{\mu\nu}F_{\mu\nu}}{2\beta^2}}\right).
\end{eqnarray}
In the low energy effective action of heterotic string theory, $\alpha$ is proportional to the inverse
string tension with positive coefficient [25]. Thus in this paper we focus on the case with a
positive Gauss-Bonnet coefficient $\alpha>0$.

In the limit $D\to 4$, the spherically symmetric solution takes the following form  \cite{Yang:2020jno}
\begin{eqnarray}
ds^2&=&-f(r)dt^2+\frac{1}{f(r)}dr^2+r^2d\Omega_2,\nonumber\\
f(r)&=&1+\frac{r^2}{2\alpha}\Big[1-\left(1+4\alpha\left(\frac{2M}{r^3}-\frac{1}{l^2}
-\frac{2\beta^2}{3}+\frac{2\beta^2}{3}\sqrt{1+\frac{Q^2}{\beta^2r^4}}\right.\right.\nonumber\\
&&\left.\left.-\frac{4Q^2}{3r^4}~_{2}F_1[\frac{1}{4},\frac{1}{2},\frac{5}{4},-\frac{Q^2}{\beta^2 r^4}]\right)\right)^{1/2}\Big],\label{solution}
\end{eqnarray}
where $M$ and $Q$ are the mass and charge of BI AdS black hole respectively, and $_{2}F_1$
is the hypergeometric function.
The electromagnetic potential difference $(\Phi)$ between the horizon and infinity is
\begin{eqnarray}
\Phi={\frac{Q}{r_+}}~_{2}F_1[\frac{1}{4},\frac{1}{2},\frac{5}{4},-\frac{Q^2}{\beta^2 r_+^4}].
\end{eqnarray}
In the limit of $\beta\rightarrow\infty$, $f(r)$ recovers the Maxwell
charged RN-AdS-like black hole solution in the 4D EGB gravity\cite{Fernandes:2020rpa}
\begin{eqnarray}
f(r)=1+\frac{r^2}{2\alpha}\Big[1-\sqrt{1+4\alpha\left(\frac{2M}{r^3}
-\frac{1}{l^2}-\frac{Q^2}{r^4}\right)}\Big].
\end{eqnarray}
When the GB term is switched off, i.e., $\alpha=0$, $f(r)$
reduces to the BI AdS black hole solution obtained in the
Einstein-Born-Infeld(EBI) gravity\cite{Dey:2004yt,Cai:2004eh}.

In terms of the horizon radius $r_+$, the mass $M$, Hawking temperature $T$ and entropy $S$
of four dimensional BI AdS black hole can be written as
\begin{eqnarray}
&&M=\frac{r_+}{2}\Big[1+\frac{r_+^2}{l^2}+\frac{\alpha}{r_+^2}
+\frac{2\beta^2r_+^2}{3}\left(1-\sqrt{1+\frac{Q^2}{\beta^2r_+^4}}\right)
+\frac{4Q^2}{3r_+^2}~_{2}F_1[\frac{1}{4},\frac{1}{2},\frac{5}{4},-\frac{Q^2}{\beta^2 r^4}]\Big],\label{M}\\
&&T=\frac{1}{4\pi r_+l^2(r_+^2+2\alpha)}\Big[3r_+^4+l^2\left(r_+^2-\alpha +2\beta^2r_+^4\left(1-\sqrt{1+\frac{Q^2}{\beta^2r_+^4}}\right)\right)\Big],\label{T}\\
&&S=\pi r_+^2+4\pi\alpha\ln{\frac{r_+}{\alpha}}.\label{S}
\end{eqnarray}

In the extended phase space, the cosmological constant $\Lambda$ is regarded as a
variable and also identified with the thermodynamic pressure $P=-\frac{\Lambda}{8\pi}$
in the geometric units $G_N=\hbar=c=k=1$. Then, the black hole mass $M$ is considered
as the enthalpy $H$ rather than the internal energy of the gravitational system.
The corresponding thermodynamical volume $V$ is given by
\begin{eqnarray}
V=\left(\frac{\partial M}{\partial P}\right)_{S,Q,\alpha,\beta}=\frac{4\pi r_+^3}{3}.
\end{eqnarray}
With all the above thermodynamic quantities at
hand, it is easily verified that the first law of black hole
\begin{eqnarray}
dH=TdS-VdP+{\cal A}d\alpha+\Phi dQ-{\cal B}d\beta,
\end{eqnarray}
where ${\cal B}$ is BI vacuum polarization and  ${\cal A}$ is the conjugate quantity of GB
coupling parameter $\alpha$ \cite{Yang:2020jno}.

The behavior of Gibbs free energy $G$ is important to determine the thermodynamic phase
transition. The free energy $G$ obeys the following thermodynamic relation $G=H-TS$ with
\begin{eqnarray}
G&=&\frac{r_+}{2}+\frac{4P\pi r_+^3}{3}+\frac{\beta^2 r_+^3}{3}\left(1-\sqrt{1+\frac{Q^2}{\beta^2r_+^4}}\right)+
\frac{\alpha}{2r_+}+\frac{2Q^2}{3r_+}~_{2}F_1[\frac{1}{4},\frac{1}{2},\frac{5}{4},-\frac{Q^2}{\beta^2 r^4}]\nonumber\\
&&-\frac{3(r_+^2+4\alpha\ln{\frac{r_+}{\alpha}})}{2r_+(r_+^2+2\alpha)}\left(r_+^2-\alpha+8\pi P+2\beta^2r_+^4(1-\sqrt{1+\frac{Q^2}{\beta^2r_+^4}})\right).\label{freeG}
\end{eqnarray}

\section{Phase transition of Born-Infeld AdS black hole}
\label{3s}

From the Hawking temperature (\ref{T}), we can obtain the equation of state
\begin{eqnarray}
P=\left(\frac{\alpha}{r_+^3}+\frac{1}{2r_+}\right)T+\frac{\alpha-r_+^2}{8\pi r_+^4}+\frac{\beta^2}{4\pi}\left(\sqrt{1+\frac{Q^2}{\beta^2r_+^4}}-1\right).\label{eos}
\end{eqnarray}
The crucial information about the equation of state is encoded in the number of
critical points it admits. As usual, a critical point occurs when $P$ has an inflection point
\begin{eqnarray}
\frac{\partial P}{\partial r_+}\Big|_{T=T_c, r_+=r_c}
=\frac{\partial^2 P}{\partial r_+^2}\Big|_{T=T_c, r_+=r_c}=0.\label{inflection}
\end{eqnarray}
Then we can obtain corresponding critical values
\begin{eqnarray}
&&T_c=\frac{r_c^2-2\alpha-2Q^2(1+\frac{Q^2}{\beta^2r_c^4})^{-1/2}}{2\pi r_c(r_c^2+6\alpha)},\label{critTc}\\
&&P_c=\frac{r_c^4-5\alpha r_c^2-2\alpha^2}{8\pi r_c^4(r_c^2+6\alpha)}+\frac{\beta^2}{4\pi\sqrt{1+\frac{Q^2}{\beta^2r_c^4}}}\left(1-\sqrt{1+\frac{Q^2}{\beta^2r_c^4}}\right)
-\frac{Q^2(r_c^2+2\alpha)}{4\pi r_c^4(r_c^2+6\alpha)\sqrt{1+\frac{Q^2}{\beta^2r_c^4}}},\label{critPc}
\end{eqnarray}
where $r_c$ is determined by
\begin{eqnarray}\label{critrc}
-r_c^4+12\alpha r_c^2+12\alpha^2+\frac{2\beta r_c^2Q^2\left(Q^2(r_c^2-6\alpha)
+3\beta^2r_c^4(r_c^2+2\alpha)\right)}{(Q^2+\beta^2r_c^4)^{3/2}}=0.
\end{eqnarray}
Here $P_c$, $r_c$ and $T_c$ are all positive in order the critical point to be physical.

Now we consider the critical behaviors of BI AdS black hole in the extended phase space.
Unfortunately, an analytic solution is not possible, since higher-order polynomials for Eq.(\ref{critrc}) are encountered.
Then, we proceed numerically: for a given $\alpha$ and $\beta$, we solve
the Eqs.(\ref{critTc})(\ref{critPc})(\ref{critrc}) for critical $r_c$, $P_c$ and $T_c$,
and then calculate the values of $P$, $T$ and $G$ using Eqs.(\ref{T}) and (\ref{freeG}), yielding a $G-T$ diagram.
Once the behavior of $G$ is known, we can further compute the corresponding phase diagram, coexistence lines,
and critical points in the $P-T$ plane.

\begin{figure}[htb]
\centering
\subfigure[]{
\includegraphics{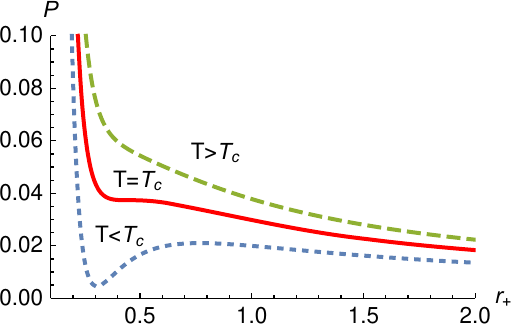}}
\hfill%
\subfigure[]{
\includegraphics{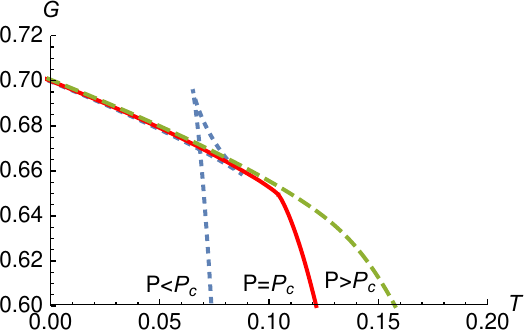}}
\hfill%
\subfigure[]{
\includegraphics{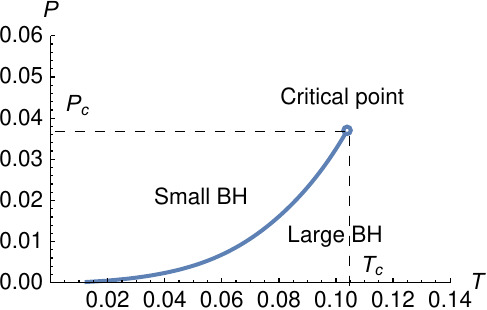}}
\caption{The $P-r_+$ isotherm diagram, Gibbs free energy $G$ (left panel) and $P-T$ diagram
for coexistence line of SBH/LBH phase transition with $\alpha=0.01$, $\beta=0.26$ and $Q=1$. }\label{fig1}
\end{figure}

In the limit of $\beta\rightarrow\infty$, Eq.(\ref{solution}) reduces to that of
charged AdS black hole \cite{Fernandes:2020rpa}, where it recovered the existence
of Van der Waals like SBH/LBH phase transition in the extended phase space \cite{Hegde:2020xlv}.
On the other hand, in the limit of $\alpha\rightarrow0$, Eq.(\ref{solution})
reduces to Born-Infeld-AdS black hole solution in the EBI gravity.
Refs.\cite{Gunasekaran:2012dq,Zou:2013owa} also display Van der Waals like
SBH/LBH phase transition for $\beta>\frac{\sqrt{3+2\sqrt{3}}}{6Q}\approx\frac{0.4237}{Q}$,
an interesting reentrant phase transition in the special region of $\frac{1}{\sqrt{8}Q}<\beta<\frac{\sqrt{3+2\sqrt{3}}}{6Q}$
and the disappearance of phase transition if $\beta<\frac{1}{\sqrt{8}Q}\approx\frac{0.35355}{Q}$.
Inspiring by these rich phase transition structures for BI AdS black hole in the EBI gravity,
here we firstly consider the small contributions of Gauss-Bonnet term (i.e., $\alpha=0.01$)
to BI AdS black hole in the 4D EGB gravity. Moreover, we assume $Q=1$ for simplification in future.
Then we find that there always exists one physical critical point when $\beta\in(0,0.339]$,
even though the BI coupling parameter $\beta$ is less than $\frac{1}{\sqrt{8}}$.
Taking $\beta=0.26(<\frac{1}{\sqrt{8}})$ for example,
we obtain the critical horizon $r_c=0.4459$, critical temperature $T_c=0.1042$ and critical pressure $P_c=0.0374$.
The $P-r_+$ isotherm diagram is plotted in Fig.\ref{fig1}(a).
The dotted line corresponds to the ``idea gas'' phase behavior when $T>T_c$,
and the Van der Waals like SBH/LBH phase transition appears in the system when $T<T_c$.
The behavior of the Gibbs free energy $G$ is important to
determine the thermodynamic phase transition. We see that the $G$ surface
in Fig.\ref{fig1}(b) demonstrates the characteristic ``swallow tail'' behavior,
which shows the Van der Waals like SBH/LBH phase transition in the system for $T<T_c$.
The coexistence line in the ($P,T$) plane by finding a curve where the Gibbs free
energy and temperature coincide for small and large black holes.
The coexistence line is very similar to that in the Van der Waals fluid.
The critical point is shown by a small circle at the end of the coexistence line, see Fig.\ref{fig1}(c).

\begin{figure}[htb]
\centering
\subfigure[]{
\includegraphics{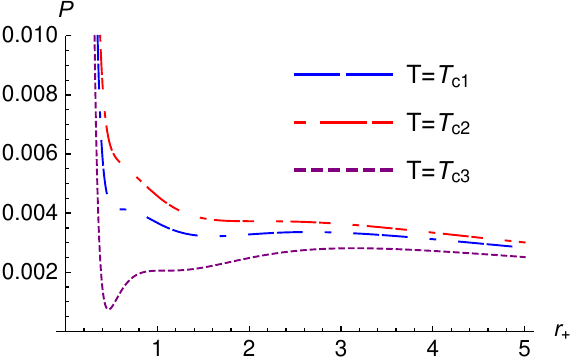}}
\hfill%
\subfigure[]{
\includegraphics{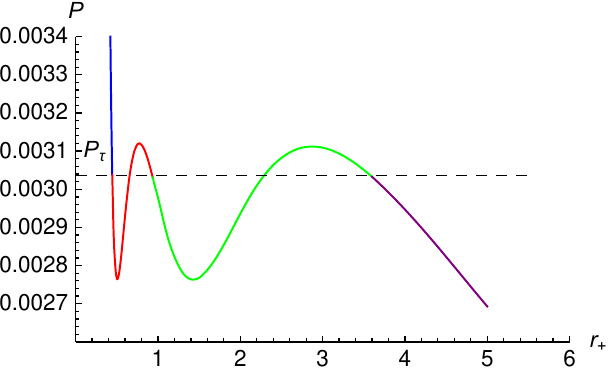}}
\caption{The $P-r_+$ diagram with $\alpha=0.01$, $\beta=0.38$ and $Q=1$.
Left: Here the (critical) temperatures are $T_{c1}=0.0435$, $T_{c2}=0.0453$ and $T_{c3}=0.0404$.
Right: The triple point in the $P-r_+$ diagram corresponds to an isotherm $(T_{\tau}=0.0422)$
for which a `double' Maxwell equal area holds for the same pressure $P_\tau=0.00304$.}\label{fig2}
\end{figure}

\begin{figure}[htb]
\centering
\subfigure[$P=0.0045$]{
\includegraphics[width=0.3\textwidth]{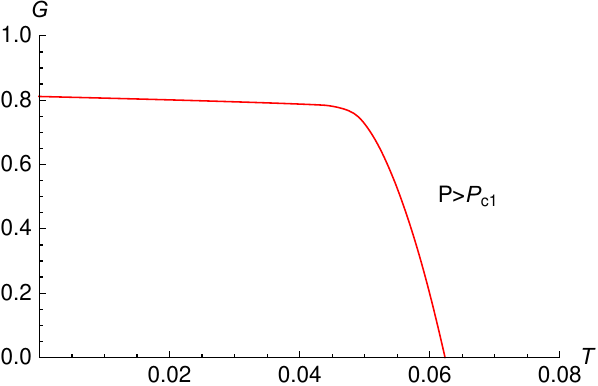}}
\hfill%
\subfigure[$P=0.004$]{
\includegraphics[width=0.3\textwidth]{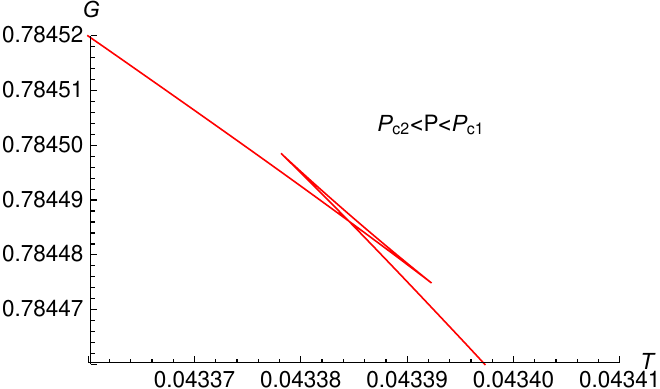}}
\hfill%
\subfigure[$P=0.0035$]{
\includegraphics[width=0.3\textwidth]{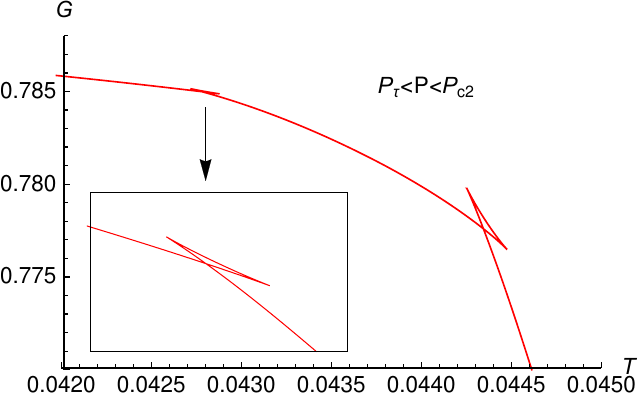}}
\hfill%
\subfigure[$P=P_{\tau}=0.00304$]{
\includegraphics[width=0.3\textwidth]{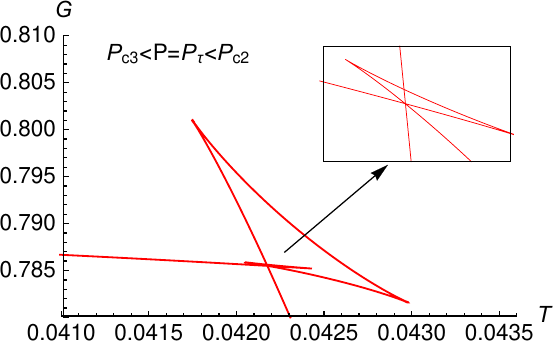}}
\hfill%
\subfigure[$P=0.0028$]{
\includegraphics[width=0.3\textwidth]{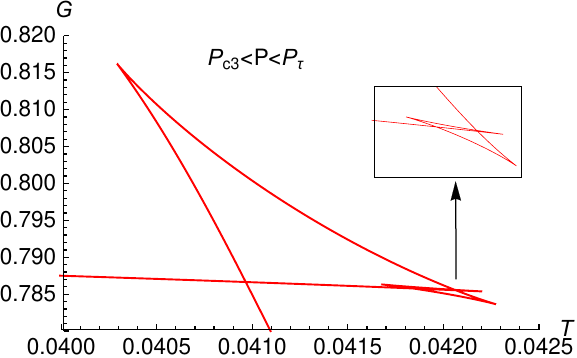}}
\hfill%
\subfigure[$P=0.0018$]{
\includegraphics[width=0.3\textwidth]{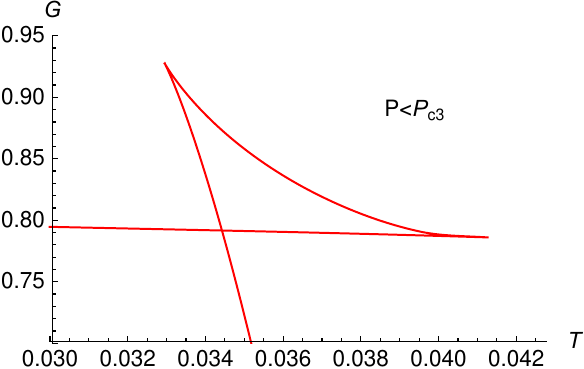}}
\caption{The Gibbs free energy $G$ as a function of $T$ for different pressures and
fixed $\alpha=0.01$, $\beta=0.38$ and $Q=1$. }\label{fig3}
\end{figure}

In the range of $\beta\in[0.34,0.4)$, Eq.(\ref{critrc}) admits three physical critical points
with positive $P_c$, $r_c$ and $T_c$. Taking $\beta=0.38$ for instance, we obtain the values of
three critical points $(P_{c1}=0.00413,T_{c1}=0.0435)$,
$(P_{c2}=0.00373,T_{c2}=0.0453)$ and $(P_{c3}=0.00206,T_{c3}=0.0404)$,
and the corresponding $P-r_+$ isotherm diagram is plotted in Fig.\ref{fig2}(a) with
different temperature $T=$0.0435, 0.0453 and 0.0404 (three critical temperatures).
In particular, there exists a special isotherm curve ($T=T_\tau$), which
denotes two Van der Waals oscillations such that the two equal area laws saturate
for the same tricritical pressure $P_\tau$, see Fig.\ref{fig2}(b).
The corresponding Gibbs free energy is illustrated in Fig.\ref{fig3}.
When $P>P_{c1}$, there is no such behavior,
and no phase transition occurs. In the ranges $P_{c2}<P<P_{c1}$,
it displays one characteristic swallow tail behavior in Fig.\ref{fig3}(b).
On the other hand, the situation becomes more subtle in the range $P_{c3}<P<P_{c2}$,
see Figs.\ref{fig3}(c)-\ref{fig3}(e). For fixed $P$ satisfying $P_{\tau}<P<P_{c2}$, we
observe two first-order small/intermediate and intermediate/large black hole phase
transitions. As the pressure $P$ is decreased, three black hole
phases (i.e., small, large and intermediate black holes)
eventually merge in a triple point. This triple point for example occurs for
\begin{eqnarray}\label{beta1}
\alpha=0.01, \quad \beta=0.38,\quad T_{\tau}=0.0422,\quad P_\tau=0.00304,
\end{eqnarray}
where small $(r_{1+}=0.44498)$, intermediate $(r_{2+}=0.93029)$ and large $(r_{3+}=3.57146)$.
The situation is a standard small/intermediate/large black hole (SBH/IBH/LBH) phase transition,
in some sense is reminiscent of a solid/liquid/gas phase transition \cite{Altamirano:2014tva}.
For fixed $P$ satisfying $P<P_{\tau}$, there always exists the SBH/LBH phase transition,
which is reminiscent of the liquid/gas phase transition of Van der Waals fluid,
see Figs.\ref{fig3}(e) and \ref{fig3}(f).

\begin{figure}[htb]
\centering
\includegraphics{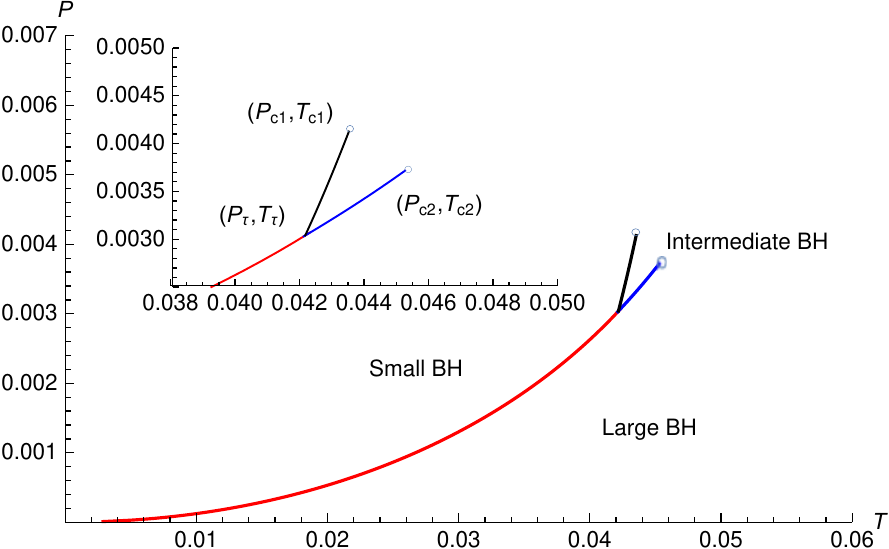}
\caption{The $P-T$ diagram of BI AdS black holes with $\alpha=0.01$, $\beta=0.38$ and $Q=1$. Other value of
$\beta\in[0.34,0.4)$ shares the same behavior of the phase diagram.}\label{fig4}
\end{figure}

The corresponding $P-T$ diagram is displayed in Fig. \ref{fig4}. It is clear that there is a triple point,
at which the small, large and intermediate black holes can coexist together. In the region of $P_\tau$ and $P_{c2}$,
there will be a SBH/IBH/LBH phase transition, which ends at $P_{c2}$. In the mean time, the first order SBH/IBH phase
transition emerges from $P_\tau$ and terminates at critical $P_{c1}$. In addition,
the system below this pressure $P_\tau$ will undergo a Van der Waals like SBH/LBH phase transition
of first order with the decreasing of $T$. Notice that the critical point $(P_{c3}, T_{c3})$
is absent from this phase diagram, since the first critical point measures the appearance of the second swallow tail,
which does not participate in phase transition.

\begin{figure}[htb]
\centering
\subfigure[]{
\includegraphics{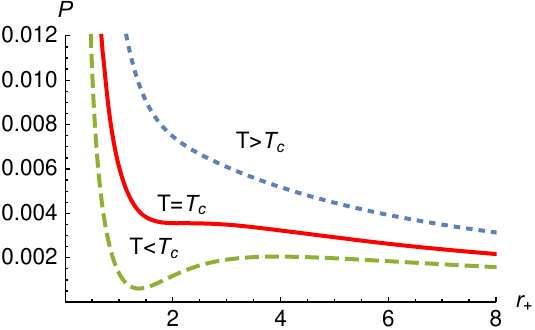}}
\hfill%
\subfigure[]{
\includegraphics{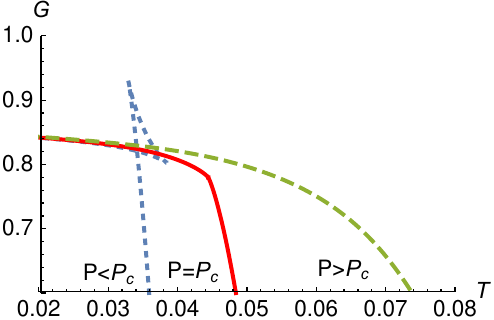}}
\hfill%
\subfigure[]{
\includegraphics{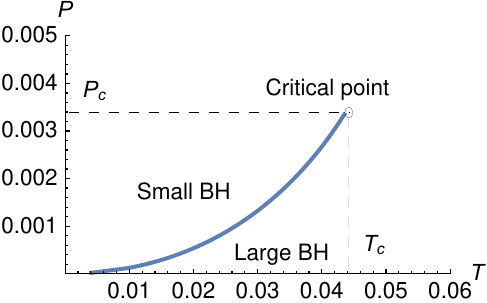}}
\caption{The $P-r_+$ isotherm diagram, Gibbs free energy $G$ (left panel) and $P-T$ diagram
for coexistence line of SBH/LBH phase transition with $\alpha=0.01$, $\beta=0.44$ and $Q=1$. }\label{fig5}
\end{figure}

For the case of $\beta\geq0.4$, we find that there always exists the Van der Waals like SBH/LBH phase transition.
The SBH/LBH phase transition occurs for $P<P_c$ and it ends at $(T_c, P_c)$, where the first-order phase
transition becomes the second-order one. The Gibbs free energy and the phase diagram are shown in Fig.\ref{fig5}.

\begin{figure}[htb]
\centering
\includegraphics{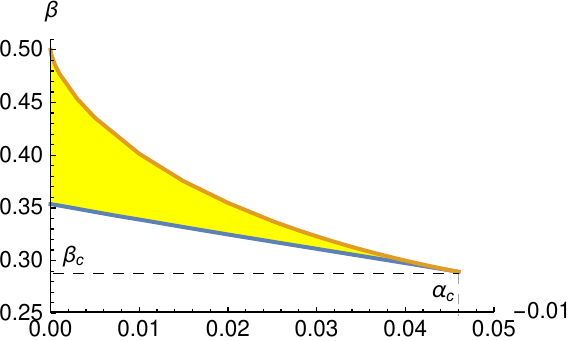}
\caption{ The available parameter region for the SBH/IBH/LBH phase transition of BI AdS black hole with $Q=1$.
The allowed area disappears when $\alpha=\alpha_c=0.04601$ and $\beta=\beta_c=0.28935$. }\label{fig6}
\end{figure}

On the other hand, we can discuss more general case for BI AdS black hole in the 4D EGB gravity.
Considering different values of Gauss-Bonnet parameter $\alpha$ and BI parameter $\beta$,
we calculate the positive values of $r_c$, $P_c$ and $T_c$ from Eqs.(\ref{critTc})(\ref{critPc})(\ref{critrc}),
and find the allowed ranges of $\alpha$ and $\beta$ corresponding to the solid/liquid/gas phase diagram like
SBH/IBH/LBH phase transition. This available parameter region for parameters $\alpha$ and $\beta$
is plotted in Fig.\ref{fig6}, where the vertical axis denotes the EBI AdS black hole with $\alpha\rightarrow0$.
It's worth to note that the Refs. \cite{Gunasekaran:2012dq,Zou:2013owa} show the existence of an interesting
reentrant phase transitions (RPT), besides the usual SBH/LBH phase transitions are observed
in the region of $\frac{1}{\sqrt{8}}\leq\beta\leq\frac{1}{2}$, since the EBI AdS black hole
possesses three real roots $r_c$, but only two physical positive roots $r_c$.
Therefore, we have reason to believe that for the BI AdS black hole in 4D Gauss-Bonnet gravity,
the Gauss-Bonnet term results in three real positive roots $r_{c1}$, $r_{c2}$ and
$r_{c3}$ of Eq.(\ref{critrc}), and corresponding physical positive temperature $T_{c1},T_{c2},T_{c3}$
and pressure $P_{c1},P_{c2},P_{c3}$. Moreover, the allowed yellow area for BI parameter $\beta$ becomes smaller
with the increasing of $\alpha$, and finally disappears when $\alpha_c=0.04601$ and $\beta_c=0.28935$.

\section{Discussion and Conclusion}
\label{4s}

In this paper, we discuss the thermodynamic behaviors of Born-Infeld AdS black hole in the novel 4D EGB gravity
by treating the cosmological constant in an extended phase space. We have written out the equations of state and
examined the phase structures with positive parameter $\alpha$ of Gauss-Bonnet term and parameter $\beta$ of Born-Infeld term.
We found there exist various thermodynamic phenomena, such as Van der Waals behaviour and tripe points.
Taking $\alpha=0.01$ for example, the system exhibits a first order SBH/LBH phase transition which resembles the
Van der Waals liquid-gas phase transition in fluids in the regions of $\beta\in(0,0.339]$ and $\beta\geq0.4$.
For the region of $\beta\in[0.34,0.4)$,  the system admits a standard SBH/IBH/LBH phase transition,
which is reminiscent of a solid/liquid/gas phase transition.

We further investigate the general case for parameters $\alpha$ and $\beta$.
We found that the triple point does not always exist in that range. The triple point and the
SBH/IBH/LBH phase transition are limited in the yellow area.
Moreover, the allowed area for BI parameter $\beta$ becomes smaller
with the increasing of $\alpha$, and finally disappears when $\alpha_c=0.04601$ and $\beta_c=0.28935$.

This work is supported by the National Natural Science Foundation of China under Grant Nos.11605152,
11675139 and 51575420, and Outstanding young teacher programme from Yangzhou University.


\begin{thebibliography}{99}
{
\bibitem{Maldacena:1997re}
  J.~M.~Maldacena,
  Adv.\ Theor.\ Math.\ Phys.\  {\bf 2}, 231 (1998)  [hep-th/9711200].
\bibitem{Gubser:1998bc}
  S.~S.~Gubser, I.~R.~Klebanov and A.~M.~Polyakov,
  Phys.\ Lett.\ B {\bf 428}, 105 (1998)  [hep-th/9802109].
\bibitem{Witten:1998qj}
  E.~Witten,
  Adv.\ Theor.\ Math.\ Phys.\  {\bf 2}, 253 (1998)  [hep-th/9802150].
\bibitem{Hawking:1982dh}
  S.~W.~Hawking and D.~N.~Page,
  Commun.\ Math.\ Phys.\  {\bf 87}, 577 (1983).
\bibitem{Witten:1998zw}
  E.~Witten,
  Adv.\ Theor.\ Math.\ Phys.\  {\bf 2}, 505 (1998)  [hep-th/9803131].
\bibitem{Dolan:2011xt}
  B.~P.~Dolan,
  Class. Quant. Grav.  \textbf{28}, 235017 (2011)  [arXiv:1106.6260 [gr-qc]].
\bibitem{Dolan:2010ha}
  B.~P.~Dolan,
  Class. Quant. Grav.  \textbf{28}, 125020 (2011)  [arXiv:1008.5023 [gr-qc]].
\bibitem{Kubiznak:2012wp}
  D.~Kubiznak and R.~B.~Mann,
  JHEP {\bf 1207}, 033 (2012)  [arXiv:1205.0559 [hep-th]].

\bibitem{Gunasekaran:2012dq}
  S.~Gunasekaran, R.~B.~Mann and D.~Kubiznak,
  JHEP \textbf{1211}, 110 (2012)  [arXiv:1208.6251 [hep-th]].
\bibitem{Hendi:2012um}
  S.~H.~Hendi and M.~H.~Vahidinia,
  Phys. Rev. D \textbf{88}, 084045 (2013)  [arXiv:1212.6128 [hep-th]].
\bibitem{Hendi:2015hgg}
  S.~H.~Hendi, R.~M.~Tad, Z.~Armanfard and M.~S.~Talezadeh,
  Eur. Phys. J. C \textbf{76}, 263 (2016)
  [arXiv:1511.02761 [gr-qc]].
\bibitem{Zhao:2013oza}
  R.~Zhao, H.~-H.~Zhao, M.~-S.~Ma and L.~-C.~Zhang,
  Eur. Phys. J. C \textbf{73}, 2645 (2013)  [arXiv:1305.3725 [gr-qc]].
\bibitem{Dehghani:2014caa}
  M.~H.~Dehghani, S.~Kamrani and A.~Sheykhi,
  Phys. Rev. D \textbf{90}, 104020 (2014)
  [arXiv:1505.02386 [hep-th]].

\bibitem{Hennigar:2015esa}
  R.~A.~Hennigar, W.~G.~Brenna and R.~B.~Mann,
  JHEP \textbf{1507}, 077 (2015)
  [arXiv:1505.05517 [hep-th]].
\bibitem{Zhang:2014jfa}
  L.~C.~Zhang, M.~S.~Ma, H.~H.~Zhao and R.~Zhao,
  Eur. Phys. J. C \textbf{74}, 3052 (2014)
  [arXiv:1403.2151 [gr-qc]].
\bibitem{Rajagopal:2014ewa}
  A.~Rajagopal, D.~Kubizňák and R.~B.~Mann,
  Phys. Lett. B \textbf{737}, 277 (2014)
  [arXiv:1408.1105 [gr-qc]].
\bibitem{Altamirano:2014tva}
  N.~Altamirano, D.~Kubiznak, R.~B.~Mann and Z.~Sherkatghanad,
  Galaxies \textbf{2}, 89 (2014)
  [arXiv:1401.2586 [hep-th]].
\bibitem{Sadeghi:2016dvc}
  J.~Sadeghi, B.~Pourhassan and M.~Rostami,
  Phys. Rev. D \textbf{94}, 064006 (2016)
  [arXiv:1605.03458 [gr-qc]].
\bibitem{Cheng:2016bpx}
  P.~Cheng, S.~W.~Wei and Y.~X.~Liu,
  Phys. Rev. D \textbf{94}, 024025 (2016)
  [arXiv:1603.08694 [gr-qc]].
\bibitem{Wei:2015ana}
  S.~W.~Wei, P.~Cheng and Y.~X.~Liu,
  Phys. Rev. D \textbf{93}, 084015 (2016)
  [arXiv:1510.00085 [gr-qc]].
\bibitem{Hendi:2015cqz}
  S.~H.~Hendi, S.~Panahiyan, B.~E.~Panah and Z.~Armanfard,
  Eur. Phys. J. C \textbf{76}, 396 (2016)
  [arXiv:1511.00598 [gr-qc]].
\bibitem{Mo:2016ndm}
  J.~X.~Mo, G.~Q.~Li and X.~B.~Xu,
  Eur. Phys. J. C \textbf{76},  545 (2016)
  [arXiv:1609.06422 [gr-qc]].
\bibitem{Xu:2014kwa}
  W.~Xu and L.~Zhao,
  Phys. Lett. B \textbf{736}, 214 (2014)
  [arXiv:1405.7665 [gr-qc]].
\bibitem{Boulware:1985wk}
D.~G.~Boulware and S.~Deser,
Phys. Rev. Lett. \textbf{55} (1985), 2656
doi:10.1103/PhysRevLett.55.2656
\bibitem{Tomozawa:2011gp}
Y.~Tomozawa,
[arXiv:1107.1424 [gr-qc]].
\bibitem{Cognola:2013fva}
G.~Cognola, R.~Myrzakulov, L.~Sebastiani and S.~Zerbini,
Phys. Rev. D \textbf{88} (2013) no.2, 024006
doi:10.1103/PhysRevD.88.024006
[arXiv:1304.1878 [gr-qc]].
\bibitem{Glavan:2019inb}
D.~Glavan and C.~Lin,
Phys. Rev. Lett. \textbf{124}, no.8, 081301 (2020)
doi:10.1103/PhysRevLett.124.081301
[arXiv:1905.03601 [gr-qc]].
\bibitem{Fernandes:2020rpa}
P.~G.~S.~Fernandes,
Phys. Lett. B \textbf{805} (2020), 135468
doi:10.1016/j.physletb.2020.135468
[arXiv:2003.05491 [gr-qc]].
\bibitem{Kumar:2020xvu}
A.~Kumar and S.~G.~Ghosh,
[arXiv:2004.01131 [gr-qc]].
\bibitem{Kumar:2020uyz}
A.~Kumar and R.~Kumar,
[arXiv:2003.13104 [gr-qc]].
\bibitem{Hegde:2020xlv}
K.~Hegde, A.~Naveena Kumara, C.~L.~A.~Rizwan, A.~K.~M. and M.~S.~Ali,
[arXiv:2003.08778 [gr-qc]].
\bibitem{Hegde:2020yrd}
K.~Hegde, A.~Naveena Kumara, C.~L.~A.~Rizwan, M.~S.~Ali and A.~K.~M,
[arXiv:2007.10259 [gr-qc]].
\bibitem{Wei:2020poh}
S.~W.~Wei and Y.~X.~Liu,
Phys. Rev. D \textbf{101} (2020) no.10, 104018
doi:10.1103/PhysRevD.101.104018
[arXiv:2003.14275 [gr-qc]].
\bibitem{HosseiniMansoori:2020yfj}
S.~A.~Hosseini Mansoori,
[arXiv:2003.13382 [gr-qc]].
\bibitem{Singh:2020mty}
D.~V.~Singh, R.~Kumar, S.~G.~Ghosh and S.~D.~Maharaj,
[arXiv:2006.00594 [gr-qc]].
\bibitem{Singh:2020xju}
D.~V.~Singh and S.~Siwach,
Phys. Lett. B \textbf{808} (2020), 135658
doi:10.1016/j.physletb.2020.135658
[arXiv:2003.11754 [gr-qc]].
\bibitem{Born:1934gh}
M.~Born and L.~Infeld,
Proc. Roy. Soc. Lond. A \textbf{144} (1934) no.852, 425-451
doi:10.1098/rspa.1934.0059
\bibitem{Hoffmann:1935ty}
B.~Hoffmann,
Phys. Rev. \textbf{47} (1935) no.11, 877-880
doi:10.1103/PhysRev.47.877
\bibitem{Dey:2004yt}
T.~K.~Dey,
Phys. Lett. B \textbf{595} (2004), 484-490
doi:10.1016/j.physletb.2004.06.047
[arXiv:hep-th/0406169 [hep-th]].
\bibitem{Cai:2004eh}
R.~G.~Cai, D.~W.~Pang and A.~Wang,
Phys. Rev. D \textbf{70} (2004), 124034
doi:10.1103/PhysRevD.70.124034
[arXiv:hep-th/0410158 [hep-th]].
\bibitem{Zou:2013owa}
D.~C.~Zou, S.~J.~Zhang and B.~Wang,
Phys. Rev. D \textbf{89} (2014) no.4, 044002
doi:10.1103/PhysRevD.89.044002
[arXiv:1311.7299 [hep-th]].
\bibitem{Yang:2020jno}
  K.~Yang, B.~M.~Gu, S.~W.~Wei and Y.~X.~Liu,
  Eur.\ Phys.\ J.\ C {\bf 80}, no. 7, 662 (2020)
  doi:10.1140/epjc/s10052-020-8246-6
  [arXiv:2004.14468 [gr-qc]].





 }\
\end{thebibliography}
\end{document}